\documentclass[11pt]{iopart}

\usepackage[super,sort&compress,comma]{natbib}
\usepackage{xcolor}
\usepackage{amsfonts}
\expandafter\let\csname equation*\endcsname\relax
\expandafter\let\csname endequation*\endcsname\relax
\usepackage{amsmath}
\usepackage{amssymb}
\usepackage{graphicx}
\usepackage{caption}
\usepackage{makecell}
\usepackage{soul}
\usepackage[colorlinks=true,linkcolor=blue,citecolor=blue]{hyperref}

\begin{document}

\title[]{From Prediction to Experimental Realization of Ferroelectric Wurtzite Al$_{1-x}$Gd$_{x}$N Alloys}

\author{Cheng-Wei Lee,\textit{$^{a}$} Rebecca W. Smaha,\textit{$^{b}$} Geoff L. Brennecka,\textit{$^{a}$} Nancy Haegel,\textit{$^{b}$} Prashun Gorai,\textit{$^{a,b,c\ast}$} Keisuke Yazawa\textit{$^{a,b\ast}$}}

\address{$^{a}$Colorado School of Mines, Golden, CO 80401. $^{b}$National Renewable Energy Laboratory, CO 80401. $^{c}$Rensselaer Polytechnic Institute, Troy, New York 12180.}
\ead{goraip@rpi.edu, Keisuke.Yazawa@nrel.gov}

\begin{abstract}
AlN-based alloys find widespread application in high-power microelectronics, optoelectronics, and electromechanics. The realization of ferroelectricity in wurtzite AlN-based heterostructural alloys has opened up the possibility of directly integrating ferroelectrics with conventional microelectronics based on tetrahedral semiconductors such as Si, SiC and III-Vs, enabling compute-in-memory architectures, high-density data storage, and more. The discovery of AlN-based wurtzite ferroelectrics has been driven to date by chemical intuition and empirical explorations. Here, we demonstrate the computationally-guided discovery and experimental demonstration of new ferroelectric wurtzite Al$_{1-x}$Gd$_x$N  alloys. First-principles calculations indicate that the minimum energy pathway for switching changes from a collective to an individual switching process with a lower overall energy barrier, at a rare-earth fraction $x$ of $x>$ 0.10--0.15. Experimentally, ferroelectric switching is observed at room temperature in Al$_{1-x}$Gd$_x$N films with $x>$ 0.12, which strongly supports the switching mechanisms in wurtzite ferroelectrics proposed previously (Lee et al., \textit{Science Advances} 10, eadl0848, 2024). This is also the first demonstration of ferroelectricity in an AlN-based alloy with a magnetic rare-earth element, which could pave the way for additional functionalities such as multiferroicity and opto-ferroelectricity in this exciting class of AlN-based materials.  
\end{abstract}

\section{Introduction}
Ferroelectric (FE) materials are potentially useful for energy-efficient computing architectures, including compute-in-memory and neuromorphic computing, and high-density non-volatile memory, among other circuit components.\cite{Ielmini2018,Oh2019} Wurtzite nitride ferroelectrics are of particular interest due to their chemical, structural, and process compatibility with commercial semiconductors such as Si, SiC and GaN, all of which exhibit tetrahedron-based crystal structures.\cite{Kim_nature_nanotech_2023} In addition, the relatively low deposition temperatures ($<$400 $^\circ$C) required for growth of wurtzite nitrides makes them compatible with back end of line (BEOL) semiconductor manufacturing processes if full integration is not necessary. Wurtzite nitrides are also generally resistant to anion migration,\cite{Ambacher_JJAP_1998, Sterntzke_JACS_1994} which is responsible for fatigue and aging in common oxide ferroelectrics such as Pb(Zr$_x$Ti$_{1-x}$)O$_3$ (PZT) and HfO$_2$.\cite{Mueller2013,Yoon2020} Here and subsequently, the term ``wurtzite'' also includes wurtzite-derived structures.\cite{Breternitz_ACA_2021} 

However, there are outstanding challenges to employing wurtzite nitride FEs in practical applications. The electric field required to switch polarization, i.e., the coercive field ($E_c$), is typically a few MV/cm, which is dangerously close to the breakdown field ($E_{B}$) of all known wurtzite ferroelectrics. Such large $E_c$ requires FE thin films with thicknesses on the order of nanometers to meet the operating voltage specifications of 1--2 V (or less) in conventional microelectronics. With such aggressive scaling, interfaces and their associated defects become increasingly dominant, which means that successful scaling in this case would require the coercive voltage to decrease with thickness faster than any associated increase in detrimental characteristics such as leakage, trap states, etc. Lowering $E_c$ and controlling bulk and interfacial defects remain major challenges for wurtzite FEs.\cite{Lee_APL_2024} Since the initial demonstration of polarization switching in Al$_{1-x}$Sc$_x$N in 2019, efforts have primarily focused on $E_c$ reduction through strain engineering.\cite{Fichtner_JAP_2019, Wang_APL_2021, Hayden_PRMater_2021, Ferri_JAP_2021, Yazawa_JMCC_2022} While in-plane biaxial tensile strain can reduce $E_c$, it is not as effective as in oxide perovskite FEs.\cite{yazawa2022landau} First-principles calculations suggest that $E_c$ can be modestly reduced by certain defects in nitrides, including native vacancies and oxygen impurities, but these studies await experimental verification.\cite{Lee_APL_2024} 

An alternative approach to aggressive physical scaling is to discover and/or develop new materials with lower $E_c$.\cite{moriwake2014, Moriwake_APLMater_2020, Dai_SA_2023, Lee2024} Recent experimental and theoretical efforts have focused on heterostructural alloying of AlN and GaN with various nitrides of $M^{3+}$ cations, e.g., $M$ = B, Sc, Y (Figure \ref{fig:comput}a). Ferroelectricity has been reported in various nitride alloys, including Al$_{1-x}$B$_x$N,\cite{Hayden_PRMater_2021} Ga$_{1-x}$Sc$_x$N,\cite{Wang_APL_2021} and Al$_{1-x}$Y$_x$N,\cite{Wang_APL_2023} with $E_c$ of similar order. Computationally-guided searches have identified additional candidates for ferroelectricity,\cite{Moriwake_APLMater_2020,Lee2024,Dai_SA_2023} but most still await experimental confirmation. 

In this work, we demonstrate the computationally-guided discovery of new FE wurtzite Al$_{1-x}$Gd$_x$N alloys with rare-earth $M^{3+}$ = Gd. Our prior work suggested that the key mechanism by which Sc additions reduce $E_c$ of Al$_{1-x}$Sc$_x$N alloys is by disrupting the nearby Al-N bonds,\cite{Yazawa_JMCC_2022} so we chose to study large ions that should be very effective in disrupting the Al-N network with more ionic bonding. We employ first-principles density functional theory calculations in conjunction with the modern theory of polarization and solid state-nudged elastic band (SS-NEB) method to predict that polarization switching can be feasibly achieved in alloy compositions of $x$ above 0.10-0.15. We show that the switching barrier, which is related to $E_c$, is reduced when the switching mechanism changes from collective to individual.\cite{Lee_FEmechansim_2023} For wurtzite materials, previously predicted minimum energy paths between positively and negatively polarized structures show that the cation-centered tetrahedra within a unit cell can flip their polarities together (collective switching) or sequentially (individual switching).\cite{Lee_FEmechansim_2023,Lee2024} 

To experimentally test our theoretical predictions, we perform a systematic study on Al$_{1-x}$Gd$_x$N thin-film capacitors and observe room-temperature ferroelectricity at compositions $x\geq$ 0.12, with $E_c$ decreasing with increasing Gd content, both in agreement with our computational predictions.
Our work broadens the palette of functionalities in wurtzite nitrides to include potential multiferroicity and opto-ferroelectricity.\cite{Hao2011,Zhang2018} 

\section{Results and Discussions}\label{Sec:results}

\subsection{Predicted Ferroelectric Parameters of Al$_{1-x}M^{3+}_x$N Alloys}
\begin{figure}[!t]
\centering
\includegraphics[width=0.95\linewidth]{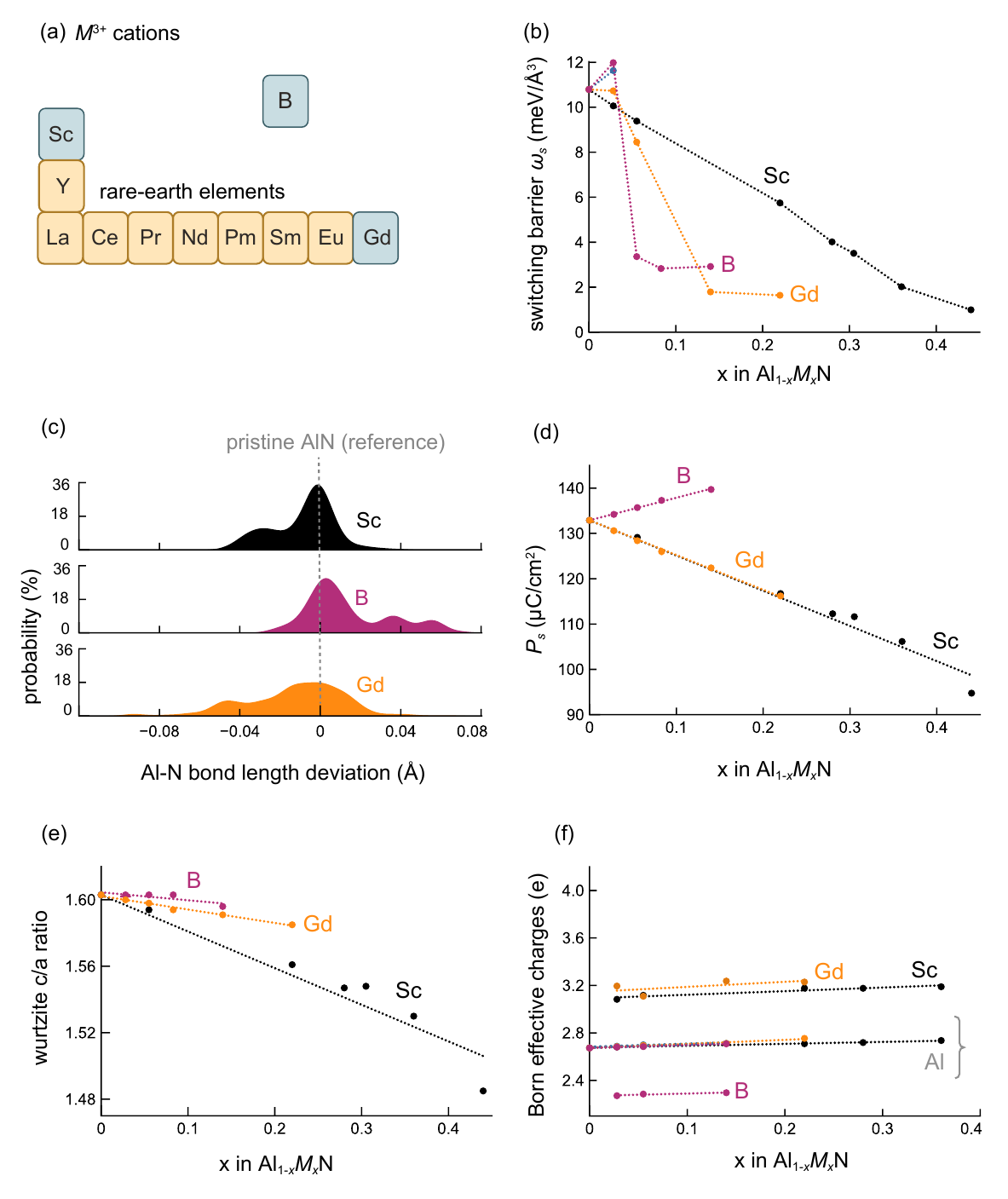}
\caption{\label{fig:comput} Ferroelectric and structural characteristics of Al$_{1-x}M_x$N alloys ($M$ = B, Sc, Gd). Data for Al$_{1-x}$Sc$_x$N are taken from Ref. \citenum{Lee_FEmechansim_2023}. (a) $M^{3+}$ cations that form $M$N nitrides in non-wurtzite ground-state structures. Alloys of AlN with elements in blue are examined in this study. (b) Calculated polarization switching barrier as a function of alloy composition. (c) Deviation in Al--N bond lengths in Al$_{0.945}M_{0.055}$N relative to pristine AlN. (d) Spontaneous polarization as a function of alloy composition. (e) Wurtzite $c/a$ ratio as a function of alloy composition. (f) Born effective charges in the polar $c$-axis direction. The standard deviations can be found in Figures S2 and S5.}
\end{figure}

FE polarization switching has been demonstrated in a few AlN-based alloys with $M^{3+}$ cations such as B, Sc, and Y. In principle, other AlN alloys (in fact, any polar material for whom $E_c < E_B$) should also exhibit ferroelectricity. We are particularly interested in rare-earth $M^{3+}$ elements because of their chemical similarity to Sc and Y. Additionally, successful demonstration of FE polarization switching with rare-earth $M^{3+}$ cations will open up opportunities to introduce additional functionalities such as optical emission, multiferroicity and/or other phenomena in wurtzite nitrides, although this is beyond the scope of the current study. We focus here on the rare-earth $M$ element Gd. We also consider AlN alloys with Sc and B for benchmarking our computational models. To model Al$_{1-x}M_x$N alloys at a given $x$, we consider four 72-atom supercells (3$\times$3$\times$2 of the wurtzite AlN unit cell) with Al randomly substituted with $M$. Since alloys are ensembles of different local environments, we consider four different alloy structures for each composition to achieve a balance between ensemble sampling and computational cost. We previously adopted this approach to model polarization switching properties of Al$_{1-x}$Sc$_x$N and found good agreement with published experimental results.\cite{Lee_FEmechansim_2023} At each $x$, we statistically averaged the structural, electronic, and ferroelectric properties across the four supercell models. 

Figure\ \ref{fig:comput}(b) shows the calculated polarization switching barrier ($\omega_s$) as a function of alloy composition ($x$) of Al$_{1-x}M_x$N ($M$ = B, Sc, and Gd). Our previous results for Al$_{1-x}$Sc$_x$N alloys are included for reference.\cite{Lee_FEmechansim_2023} For Al$_{1-x}$Sc$_x$N, we found that $\omega_s$ decreases almost linearly with increasing Sc content, which is qualitatively consistent with the reported linear decrease in experimental $E_c$ with $x$.\cite{Lee_FEmechansim_2023, Fichtner_JAP_2019,Yasuoka_JAP_2020, Yazawa_JMCC_2022} More importantly, we found that the polarization switching process at the atomic scale depends on the alloy composition. At low $x$, collective switching prevails, in which the cation and anion sublattices \textit{collectively} squeeze through one another via a nonpolar hexagonal BN-like structure. Collective switching essentially causes \textit{all} cation-centered tetrahedra to ``flip'' simultaneously along the polar axis, e.g., from pointing along $c$ to -$c$. At larger Sc compositions ($x \gtrsim$ 0.28), ``individual'' switching is observed, in which the cation-centered tetrahedra flip along the polar axis in a sequential manner rather than collectively. The collective and individual mechanisms are discussed in more detail in Ref. \citenum{Lee_FEmechansim_2023}. Of particular importance to this work is the observation that when individual switching is observed, it represents a lower energy barrier to polarization reversal than the corresponding collective switching pathway. 

For benchmarking, we calculated the polarization switching properties, including $\omega_s$ and spontaneous polarization ($P_s$), of Al$_{1-x}$B$_x$N alloys (Figure \ref{fig:comput}). The transition from collective to individual switching mechanism with increasing B composition is similar to that found in Al$_{1-x}$Sc$_x$N. At low B compositions (e.g., $x$ = 0.028), collective switching is observed in all four alloy supercells. Interestingly, $\omega_s$ increases slightly relative to AlN, which is attributed to the slightly larger kinetic barrier near the peak of the minimum energy pathway associated with moving B across the hexagonal BN plane (Figure S1). With further increase in B composition beyond $x$ = 0.055, individual switching is observed in all four alloy supercells along with a concomitant decrease in $\omega_s$, again similar to Al$_{1-x}$Sc$_x$N.\cite{Lee_FEmechansim_2023} In contrast, the transition in the switching mechanism occurs at much lower B compositions ($x$ $\sim$ 0.055) compared to Al$_{1-x}$Sc$_x$N ($x$ $\sim$ 0.28). Beyond $x$ = 0.055, $\omega_s$ plateaus, unlike in Al$_{1-x}$Sc$_x$N where $\omega_s$ progressively decreases almost linearly up to $x$ = 0.44 (highest $x$ considered). Our computational results are qualitatively in agreement with the experimental report by Hayden et al.,\cite{Hayden_PRMater_2021} in which the authors demonstrated ferroelectric behavior at $x \gtrsim$ 0.02 and plateauing of the coercive field at $x \gtrsim$ 0.13. However, we must note that even though we predicted lower $\omega_s$ for Al$_{1-x}$B$_x$N compared to Al$_{1-x}$Sc$_x$N, experimentally measured coercive fields in ferroelectric Al$_{1-x}$B$_x$N thin films are higher than in Al$_{1-x}$Sc$_x$N. It is known that experimental coercive fields are dominated by extrinsic factors such as measurement frequency,\cite{Scott1996} defects,\cite{Kitamura1998,Kim2002} and domain wall structure,\cite{Choudhury2008, Yazawa2020AFM} which are not taken into account in the DFT calculations. Future studies should address larger length-scale simulations that account for domain walls, as has been done for oxide perovskites.\cite{Shin_nature_2007}    

We find that AlN alloyed with Gd exhibits similar polarization switching behavior as alloys with Sc and B. Like B, alloying with Gd induces a slight increase in $\omega_s$ at $x$ = 0.028 followed by a drop in $\omega_s$ at $x$ = 0.055, which is associated with the transition from the collective to individual switching mechanism. However, there are subtle differences in the trends seen for B and Gd. For Gd, the initial increase in $\omega_s$ at $x$ = 0.028 and drop at $x$ = 0.55 is less abrupt compared to B. Unlike B, Gd alloying does not induce a sudden transition from collective to individual switching, but instead there is a transition region where collective switching is observed in some supercells while individual switching is seen in others, similar to Al$_{1-x}$Sc$_x$N.\cite{Lee_FEmechansim_2023} While the limited number of supercells and finite compositions studied demand caution, the results suggest that B is more effective in promoting individual switching than Gd or Sc. From these results and comparisons with Al$_{1-x}$Sc$_x$N, we predict that room temperature ferroelectric switching should be feasible in wurtzite Al$_{1-x}$Gd$_x$N for $x\geq$ 0.14.

Next, we assess the effect of alloying on spontaneous polarization ($P_s$). The calculated $P_s$ decreases with increasing Sc and Gd content ($x$) while it increases with B content in Al$_{1-x}M_x$N alloys (Figure \ref{fig:comput}d). The latter is consistent with previously reported calculations.\cite{Hayden_PRMater_2021, Liu_APL_2023} At first glance, $P_s$ increase with B content is unintuitive because the addition of B introduces large structural distortions away from the perfect wurtzite crystal structure of AlN; such distortions generally lead to lowering of $P_s$ based on our current understanding of Al$_{1-x}$Sc$_x$N alloys and the resulting reduction in wurtzite $c$/$a$ lattice parameter ratio (Figure \ref{fig:comput}e). However, $P_s$ also inversely depends on cell volume. Owing to the small size of B, alloying causes a sharp decrease in the cell volume (by $\sim$6.2\% at $x$ = 0.14), as shown in Figure S4. Therefore, the increase in $P_s$ is due to the volume contraction associated with B alloying. Alloying with Sc and Gd is associated with a quantitatively similar and almost linear decrease in $P_s$ with increasing $x$ (Figure \ref{fig:comput}e). 

A side effect of alloying AlN with $M$ = Sc and Gd is the reduction in the band gap because these $M$N nitrides are either metallic or small band-gap rocksalts.\cite{Trodahl_PRB_2007, wachter2016,Deng_AIPA_2021} h-BN has a hexagonal structure with a large indirect band gap of 5.9 eV.\cite{cassabois2016} Band gap reduction can reduce breakdown field and/or increase leakage currents -- both undesirable for dielectrics. We calculated the band gaps of Al$_{1-x}M_x$N for a limited range of compositions using hybrid DFT functional HSE06 (see Figure S3) and find that the calculated average band gap of Al$_{0.86}$Gd$_{0.14}$N ($\sim$4.5 eV) is comparable to Al$_{0.64}$Sc$_{0.36}$N ($\sim$4.6 eV). Therefore, relatively large band gaps is maintained upon alloying AlN with Gd within the range of compositions explored in this study. 

Since a transition in the switching mechanism from collective to individual switching is associated with a significant reduction in switching barrier and is thus a promising strategy for reducing $E_c$ below breakdown field, we seek to deduce the underlying composition-structure-property relationships.\cite{Lee2024, Lee_FEmechansim_2023} There is emerging evidence that larger structural distortions away from the ideal wurtzite crystal structure promote polarization switching.\cite{Lee2024} Given the large differences in the ionic radii between Al and $M$ = B, Sc, and Gd, we quantitatively investigate the local structural distortions in Al$_{1-x}M_x$N alloys. Figure \ref{fig:comput}(c) shows the distribution of deviations in Al-N bond length in Al$_{0.945}M_{0.055}$N relative to pristine AlN of the same volume (see Methods for details). The shape of the distribution is an indicator of the extent of the distortion. Small distortions will translate to tight distributions with a single strong peak near zero deviation, while larger distortions will result in wider and flatter distributions. The overall shape of the distribution in the case of Al$_{0.945}M_{0.055}$N alloys mainly reflects the distortions in the basal-plane Al--N bonds and is qualitatively different between B and rare-earths Sc and Gd. In case of B, the basal-plane Al--N bonds that connect to B via the shared N atoms are longer than the rest since B--N bonds are shorter than the average Al--N bond. The opposite behaviors are observed for Sc and Gd. The difference in $M$--N bond lengths is also related to their ground-state structures, with BN in layered hexagonal structure while ScN and GdN are in rocksalt. Among the three elements, we find that Gd alloying causes larger distortions compared to Sc and B.

In our previous studies,\cite{Yazawa_JMCC_2022, Lee_FEmechansim_2023, Lee2024} we have proposed that the increased ionicity of Sc--N bonds (relative to Al--N) promotes polarization switching in Al$_{1-x}$Sc$_x$N. $M$ are nominally 3+ cations, but $M$--N have both ionic and covalent character. Born effective charge is one indicator of the degree of iono-covalency of the bonds, but it also accounts for the lattice softness.\cite{Yazawa_JMCC_2022} Figures \ref{fig:comput}(f) shows the Born effective charges along the polar $c$ axis as a function of the alloy composition. The Born effective charges in the $a$--$b$ plane are shown in Figure S5. Gd has the highest Born effective charges followed by Sc, Al, and B, which is consistent with their Pauling electronegativities. Alloying does not significantly affect the average Born effective charges of Al, but the standard deviation increases, reflecting the local structural distortions (Figure S5).

\subsection{Experimental Verification of Ferroelectricity in Al$_{1-x}$Gd$_x$N Alloys}

\begin{figure}[!t]
\centering
\includegraphics[width=0.5\linewidth]{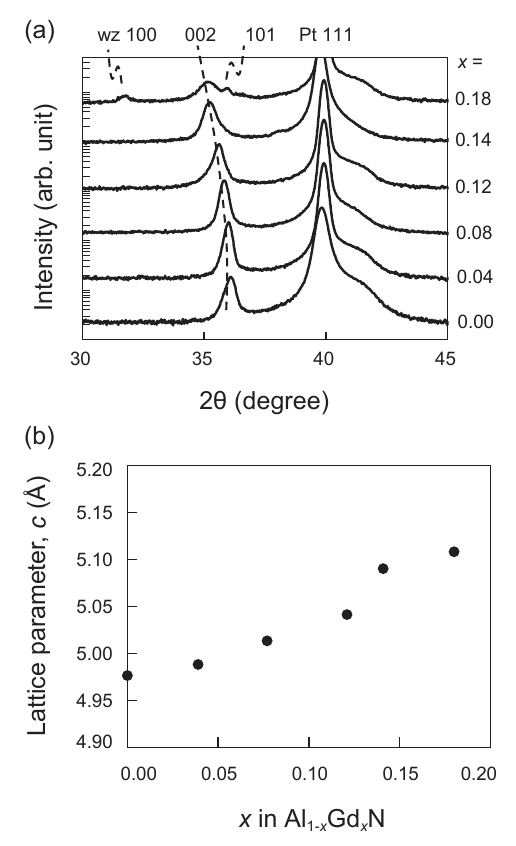}
\caption{\label{fig:xtal} Crystal structure of Al$_{1-x}$Gd$_x$N films on Pt/TiO$_2$/SiO$_2$/Si substrate. (a) XRD 2$\theta$ profiles shows pure wurtzite phase and polar texture up to $x$ = 0.14. (b) Lattice parameter increases with Gd content indicating solid solution.} 
\end{figure}

To investigate the predicted ferroelectricity of Gd-substituted AlN, we deposited Al$_{1-x}$Gd$_x$N thin films of various compositions ($x$) on platinized silicon substrates to fabricate ferroelectric capacitors (see Methods). The Al$_{1-x}$Gd$_x$N films crystallize in a phase-pure wurtzite structure up to at least $x$ = 0.18, with strong polar-axis texture up to $x$ = 0.14 (Figure \ref{fig:xtal}a). The diffraction patterns show wurtzite diffraction peaks originating from the Al$_{1-x}$Gd$_x$N films as well a (111) peak from the Pt bottom electrode. The (100) and (101) wurtzite diffraction peaks observed in the film with $x$ = 0.18 indicate various out-of-plane crystal orientations whose polar axes are not parallel to the surface normal direction and extrinsically affect the observed ferroelectric properties. Hereafter, we focus on the films with $x \le$ 0.14 for the ferroelectric measurements. Successful incorporation of Gd into AlN is confirmed by the increase in the lattice parameter with Gd content (Figure \ref{fig:xtal}b). The lattice parameter $c$ is in good agreement with our previous report on Al$_{1-x}$Gd$_x$N films on Si substrates,\cite{Smaha_ChemMat_2022} and the trend is consistent with our DFT results (Figure S4).

\begin{figure}[!t]
\centering
\includegraphics[width=0.5\linewidth]{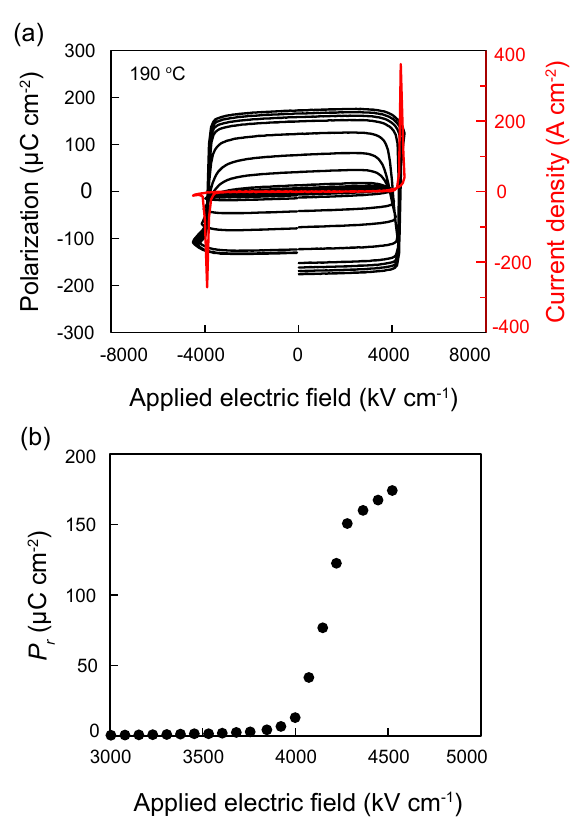}
\caption{\label{fig:P-E} Ferroelectric hysteresis loops for Al$_{0.88}$Gd$_{0.12}$N at 190 $^\circ$C. (a) Nested polarization-electric field hysteresis and current loops show ferroelectric switching. (b) Saturation behavior of remanent polarization ($P_r$).}
\end{figure}

Unambiguous ferroelectric hysteresis loops are observed for the Al$_{1-x}$Gd$_x$N films. As shown in Figure \ref{fig:P-E}a, the nested ferroelectric polarization -- electric field loops of Al$_{0.88}$Gd$_{0.12}$N taken at 190 $^\circ$C are square and possess a large remanent polarization ($>$100 $\mu$C cm$^{-2}$), comparable to other wurtzite AlN-based ferroelectrics.\cite{Fichtner_JAP_2019,Yazawa2021} Along with the hysteresis loops, distinct switching current peaks at the coercive field ($\sim$4000 kV cm$^{-1}$) are observed, supporting the polarization reorientation under the electric field. The ferroelectric switching between the two distinct up- and down-polar states is confirmed with the abrupt increase of remanent polarization at the coercive field and polarization saturation beyond the coercive field, as shown in Figure \ref{fig:P-E}(b). The gradual increase of the polarization is associated with the leakage current contribution, which complicates the quantification of the precise remanent polarization value from the plot.

\begin{figure}[!t]
\centering
\includegraphics[width=0.5\linewidth]{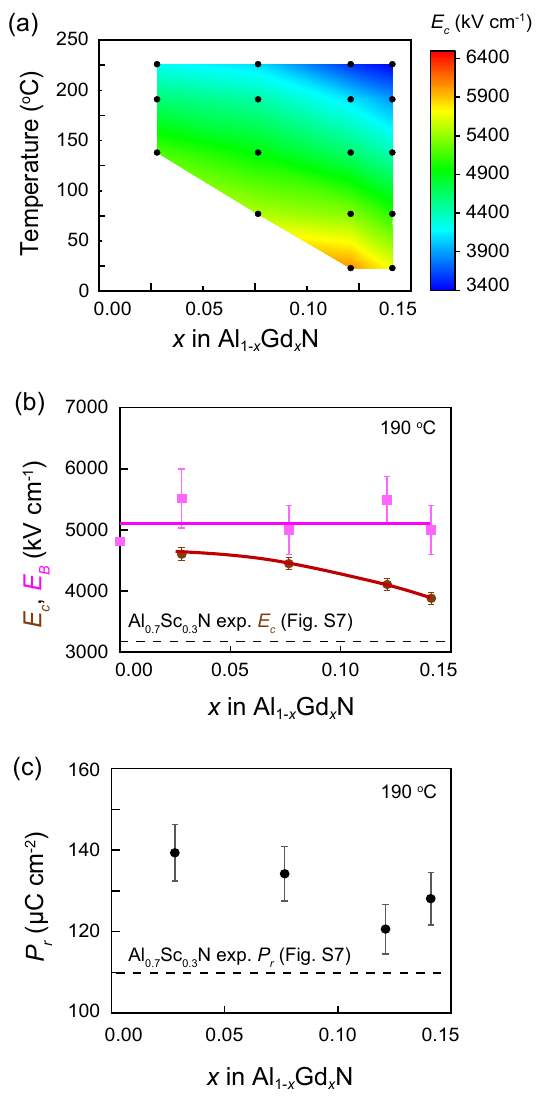}
\caption{\label{fig:FE_map} Ferroelectric properties of  Al$_{1-x}$Gd$_x$N in composition and measurement temperature space. (a) Coercive field contour map in ferroelectric switchable range. Black dots represent measurement points, and the color map is linearly interpolated. (b) Coercive field and breakdown field as a function of Gd content at 190 $^{\circ}$C, showing that the coercive field decreases with Gd content while the breakdown field does not change significantly. (c) Remanent polarization decreases with Gd content. For comparison, the coercive field of an Al$_{0.7}$Sc$_{0.3}$N film at the same measurement condition (Figure S7, Supplementary Information) is shown as a horizontal dashed line.} 
\end{figure}

To understand the effects of Gd substitution and compare the experiments to the computational predictions, we investigated the ferroelectric properties as a function of composition. Figure \ref{fig:FE_map}a illustrates the coercive field contour map for various $x$ and measurement temperatures, based on hysteresis loops in the variable space (Figure S6). At room temperature, the films with ${x \ge}$ 0.12 show ferroelectric switching, which is quantitatively consistent with the computational prediction showing an abrupt switching barrier decrease at $x$ $\approx$ 0.14 (Figure \ref{fig:comput}b). At room temperature, the films with ${x <}$ 0.12 do not show ferroelectric switching and undergo dielectric breakdown at an electric field that is lower than the coercive field. The switchable composition range widens with increased measurement temperature. Above 140 $^\circ$C, polarization switching is observed as low as $x$ = 0.03. In addition, the coercive field decreases with increased measurement temperature, which is consistent with other wurtzite FEs.\cite{Drury_micromachines_2022, Zhu_APL_2021, Hayden_PRMater_2021, Ferri_JAP_2021}

The coercive field also decreases with increased Gd content. Figure 4b shows the extracted coercive fields and breakdown fields under ferroelectric loop measurement conditions (10 kHz bipolar triangular voltage) as a function of Gd content at 190 $^\circ$C. The trend of decreasing coercive field is consistent with the calculated barrier height (Figure \ref{fig:comput}b). Note that the absolute value of the experimental coercive field is not directly transferable from the barrier height due to extrinsic effects such as defects and domain structures.\cite{Kitamura1998,Kim2002,Choudhury2008,Yazawa2020AFM}

The relationship between the coercive field and (greater) breakdown field enables ferroelectric switching. In the current study, the breakdown field does not change significantly with Gd content, although there are large error bars attributed to the inter-device variability ($\sim\pm$400 kV cm$^{-1}$). Extrapolating coercive field from this study to a composition of $x$ = 0 suggests a value of $\sim$5000 kV cm$^{-1}$, which is comparable to both the breakdown strength for all samples in the current study and the reported coercive field of AlN at 190 $^{\circ}$C\cite{Zhu_APL_2021}. Thus, the low-temperature and low-Gd content region of Fig.~\ref{fig:FE_map}a where switching is not observed is explained by the coercive field increasing beyond the breakdown field. This relationship and its temperature-dependence is shown further in Figure S8, Supplementary Information. 

Remanent polarization also decreases with Gd content. The remanent polarization values are obtained by removing leakage and capacitive contributions from switching current measurements (See Methods and Figure S9 in Supplementary Information).\cite{Yazawa2020} Figure \ref{fig:FE_map}c shows the extracted remanent polarization as a function of Gd content at 190 $^\circ$C. The uncertainty is dominated by uncertainty in defining the electrode boundary and thus device size, but the trend is comparable to the DFT calculations (Figure \ref{fig:comput}e).  

We note that Al$_{1-x}$Gd$_x$N is the first example of an AlN-based ferroelectric material that incorporates a magnetic rare earth cation and thus might exhibit interesting magnetic properties. To test whether the addition of magnetic (Gd$^{3+}$) to AlN might induce magnetic order and thus multiferroicity, we measured the magnetic susceptibility of an Al$_{0.86}$Gd$_{0.14}$N film grown on a Si substrate. As shown in Figure S10, no evidence of long-range magnetic ordering was observed down to $T=2$ K. The moment as a function of applied magnetic field exhibits no hysteresis curve at $T=2$ K, and no peak is observed as a function of temperature: the data are consistent with paramagnetism. Some of the primary challenges to introducing multiferroicity to this family of materials will be 1) introducing a sufficient amount of the magnetic cation to induce long-range magnetic order, and 2) the tetrahedral coordination may often lead to antiferromagnetic interactions instead of ferromagnetic interactions, which are more technologically useful. However, single-phase multiferroics are exceedingly rare, so if this family of materials can successfully be designed to simultaneously exhibit ferroelectricity and magnetic order (i.e., by incorporating enough of a magnetic cation), it will be not only scientifically interesting (as these properties are generally mutually exclusive) but also highly relevant to applications in energy technologies such as spintronics and low-power computing.
 
\section{Conclusion}\label{sec:conclusion}
In summary, we computationally predict and experimentally verify ferroelectricity at room temperature in Al$_{1-x}$Gd$_x$N alloys. Our calculations show that Al$_{1-x}$Gd$_x$N, along with Al$_{1-x}$B$_x$N, have a drastic change in polarization switching barrier due to a change in switching mechanisms as the amount of alloying element is increased. Further structural analysis shows that larger local distortions due to alloying promotes this change in switching mechanism from collective to individual. Our predictions for the alloy compositions where this change in switching mechanism occurs for Al$_{1-x}$B$_x$N and Al$_{1-x}$Gd$_x$N are qualitatively consistent with experimental results, with ferroelectric switching clearly demonstrated at low alloying content ($x \lesssim$ 0.1).  These results provide strong evidence that promoting individual switching at the atomic scale can assist with ferroelectric switching in wurtzite AlN-based alloys by reducing coercive fields. We note, however, that the predicted intrinsic switching barriers are not yet quantitatively comparable to experimentally measured coercive field values. Lastly, the Al$_{1-x}$Gd$_x$N alloy is the first experimentally demonstrated wurtzite ferroelectric material with a rare-earth cation. This suggests a route to possible multi-functionality such as multiferroicity and opto-ferroicity in tetrahedral-based microelectronics.

\section{Methods}

\subsection{Computational}

\subsubsection{DFT Parameters}
We used Vienna Ab-initio Simulation Package (VASP 5.4.4) to perform density functional theory (DFT) calculations.\cite{kresse1996} All the calculations are based on the following description unless specified. We used plane-wave basis set with cutoff kinetic energy of 340 eV. We used the generalized gradient approximation (GGA) of Perdew-Burke-Ernzerhof (PBE) as the exchange correlation functional.\cite{perdew1996} The valence electrons were treated with the projector-augmented wave (PAW) method.\cite{PAW} Specifically, we used the following pseudopotentials distribtued with VASP: Al 04Jan2001, N\textunderscore s 07Sep2000, Gd\textunderscore 3 06Sep2000, and B\textunderscore s 22Jan2003. We applied an effective on-site Hubbard potential $U$ = 3.0 eV to the Sc $d$ orbitals. The Brillouin zone was sampled using automatically generated $\mathrm{\Gamma}$-centered Monkhorst-Pack \textbf{\textit{k}}-point grid defined by a length factor ($R_k$) of 20.

\subsubsection{Switching Barrier}
We applied the solid-state nudged elastic band (SS-NEB) method to find the polarization switching pathways between the positive- and negative-polarity structures.\cite{Sheppard_JCP_2012} We utilized the VASP Transition State Theory (VTST) tools developed by Henkelman and Jonsson,\cite{Henkelman_JCP_2000} as implemented in the vtst-182 code. The positive- and negative-polarity structures were fully relaxed (cell volume and shape, and ion positions) and we generated the initial images along the pathway via linear interpolation between them. The SS-NEB calculations were iterated until the forces on each atom were below the tolerance of 10$^{-2}$ eV/\AA. We note that a large number of intermediate images ($>$ 100) are needed to fully resolve the complex minimum energy pathways for the alloys. More details, including guidelines for convergence, can be found in Ref. \citenum{Lee_FEmechansim_2023}.

\subsubsection{Polarization}
We applied the modern theory of polarization, as implemented in VASP 5.4.4., to calculate the electronic contribution to the spontaneous polarization.\cite{King-Smith_PRB_1993, Resta_RMP_1994} This approach is based on the Berry phase approximation and we chose (0.25, 0.25, 0.25) crystal coordinates as the center of the reference frame for dipole calculations. We assumed point charges for the ionic contribution. Using the SS-NEB pathway, we utilized the implementation in Pymatgen to determine smooth adiabatic paths and lattice quanta.\cite{Smidt_SD_2020} We visually inspected the smoothness of the paths and manually identified the smooth ones if the smooth path finding algorithm in Pymatgen failed. 

\subsubsection{Electronic Band Gap}
 We used the hybrid functional (HSE06) to fully relax the 72-atom supercells for AlN-based alloys and to calculate their electronic band structure. The Brillouin zone was sampled using a $\mathrm{\Gamma}$-centered Monkhorst-Pack $2\times2\times2$ \textbf{\textit{k}} grid. Band gaps of the alloys were obtained by averaging over four supercells. 
 
\subsubsection{Born Effective Charge}
We calculated the Born effective charges based on self-consistent response to finite electric field, with the fourth order finite difference stencil, as implemented in VASP. Homogeneous electric fields in the $x$, $y$, and $z$ directions are  set at 0.01 eV/\AA. Damped molecular dynamics method was used to calculate the self-consistent response and time step of 0.1 was chosen to ensure convergence. We used a kinetic energy cutoff of 520 eV and $\mathrm{\Gamma}$-centered Monkhorst-Pack \textbf{\textit{k}}-point grids corresponding to $n_{kpts}$ $\times$ $n_{atoms}$ $\approx$ 2000, where $n_{kpts}$ and $n_{atoms}$ are the numbers of \textbf{\textit{k}} points and atoms in the cell, respectively. 

\subsubsection{Structural Distortion}
We used the wurtzite $c/a$ lattice parameter ratio to characterize the global structural changes induced by alloying. The $c/a$ ratio is calculated for the fully-relaxed structures and we only considered the lengths of the lattice vectors $a$, $b$, and $c$, such that
\begin{equation}
    c/a = \frac{|c|}{0.5 (|a|+|b|)}
\end{equation}

\noindent
We calculated the Al--N bond lengths to capture the local structural changes due to alloying. We used the Al--N bond lengths in relaxed wurtzite AlN as the reference. The bonds are categorized into axial and basal bonds based on the direction of the polar axis. Considering the volume change due to alloying, we also scale the reference bond lengths by $\sqrt[3]{\frac{V\mathrm{_{alloy}}}{V_{\mathrm{AlN}}}}$, where $V\mathrm{_{alloy}}$ and $V_{\mathrm{AlN}}$ are volumes of relaxed supercells of AlN-based alloys and wurtzite AlN, respectively.  The Al--N bonds in relaxed structures of AlN-based alloys are also categorized into axial and basal bonds. We calculated the deviation from the reference Al--N bond lengths only within the same sets.

\subsection{Experimental}

\subsubsection{Synthesis}
The 200-260 nm Al$_{1-x}$Gd$_x$N films were deposited on platinized silicon substrates via reactive RF magnetron sputtering using the following growth conditions: 2 mTorr of Ar/N$_2$ (5/15 sccm flow), and a target power density of 7.4 W/cm$^2$ on a 2'' diameter Al target and 0-2.5 W/cm$^2$ on a 2'' diameter Gd target. The substrate was rotated and heated to 400$^{\circ}$C during deposition. The base pressure, partial oxygen and water vapor pressure at 400$^{\circ}$C were $<$ 2 × 10$^{-7}$ torr, P$_{O2}$ $<$ 2 × 10$^{-8}$ torr and P$_{H2O}$ $<$ 1 × 10$^{-7}$ torr, respectively.  Top Au (100 nm)/Ti (20 nm) contacts (50 µm in diameter) were deposited on the Al$_{1-x}$Gd$_x$N film via electron beam evaporation through a metal shadow mask pattern. 

\subsubsection{Characterization}
The crystal structure of the film is investigated using X-ray diffraction (XRD) on Bruker D8 Discover diffractometers. The XRD profiles shown in Fig. X was obtained by integration of 2-dimensional detector images. The c lattice parameter is determined by pseudo-Voigt fitting of the wurtzite (002) diffraction peak assuming a hexagonal lattice. The cation composition is measured using X-ray fluorescence analysis calibrated with microprobe analysis described in our previous study\cite{Smaha_ChemMat_2022}.

Ferroelectric polarization – electric field hysteresis and current – electric field curve measurements were taken with a Precision Multiferroic system from Radiant Technologies. The applied triangle excitation field was up to 6.0 MV cm$^{-1}$ at 10 kHz. To wake-up the polarization-electric field hysteresis, we applied the electric field $\sim$5 times sequentially. The remanent polarization, P$_r$, was calculated from time dependent current under the triangular excitation field, which is expressed as
\begin{equation}
    P_r = \frac{1}{2}\int i_s dt = \frac{1}{2}\int (i-i_c-i_l) dt
\end{equation}
where $j_s$ is the switching current density, $j$ is the measured current density, $j_c$ is the capacitive current density, and $j_l$ is the leakage current density. The capacitive current density is a positive or negative offset under the triangular electric field determined by the electric field slope and capacitance. The leakage current density is estimated assuming that there is no leakage hysteresis, namely the leakage current density is identical at the same electric field regardless of ramping up or down (Figure S9). 

Magnetic properties were measured via superconducting quantum interference device (SQUID) magnetometry in a Quantum Design Magnetic Properties Measurement System (MPMS3) with the Vibrating Sample Magnetometer. The films were measured from $T=$2--350 K under applied fields from -7 to +7 T. The measured Al$_{0.86}$Gd$_{0.14}$N film was an approximately 5 $\times$ 5 mm piece grown on a pSi substrate. It was approximately 250 nm thick. To isolate the signal of the film, a bare substrate was also measured and subtracted. 

\section*{CRediT Statement}
\textbf{Cheng-Wei Lee}: Investigation (primary), Data Curation (primary), Writing (original draft and editing).
\textbf{Rebecca W. Smaha}: Investigation, Writing (Editing).
\textbf{Geoff L. Brennecka}: Writing (Editing), Supervision, Project Administration.
\textbf{Nancy Haegel}: Writing (Editing), Supervision, Project Administration. 
\textbf{Prashun Gorai}: Conceptualization, Investigation, Data Curation, Writing (Editing), Supervision.
\textbf{Keisuke Yazawa}: Conceptualization, Investigation (primary), Data Curation (primary), Writing (original draft and editing), Supervision.

\section*{Acknowledgements}
This work was authored in part by the National Renewable Energy Laboratory, operated by Alliance for Sustainable Energy, LLC, for the U.S. Department of Energy (DOE) under Contract No. DE-AC36-08GO28308. Funding provided by the Department of Energy Basic Energy Sciences (BES), with additional support from Advanced Scientific Computing Research (ASCR), under program ERW6548. The work was supported by the National Science Foundation under Grant No. DMR-2119281. This work  used computational resources sponsored by the Department of Energy's Office of Energy Efficiency and Renewable Energy, located at NREL. The authors also express their appreciation to Dr Andriy Zakutayev for feedback to an early version of this manuscript. The views expressed in the article do not necessarily represent the views of the DOE or the U.S. Government.\\

\section*{Conflict of Interest} 
They authors declare no conflict of interest

\section*{Data Availability Statement} 
Data associated with this study are available from the corresponding authors upon request.

\clearpage
\bibliography{AlGdN_FE}

\end{document}


\renewcommand{\thetable}{S\arabic{table}}
\renewcommand{\thefigure}{S\arabic{figure}}

\clearpage
\section{Polarization Pathways for Al$_{0.972}M_{0.028}$N Alloys ($M$ = Sc, B, Gd)}

Figure S1(a) shows the minimum energy paths (MEP), determined by the solid-state nudged elastic band (SS-NEB) method, for  Al$_{0.972}M_{0.028}$N alloys ($M$ = Sc, B, Gd). The polarization was calculated by the modern theory of polarization using the MEPs, which for AlN and all the Al$_{0.972}M_{0.028}$N alloys except for Al$_{0.972}$B$_{0.028}$N follow the typical shape for collective switching. In contrast, the Al$_{0.972}$B$_{0.028}$N alloy has a distinctive bump at the peak of its MEP. Structures 1 and 2 in Figure S1(b) are the associated structures before and after the bump, respectively. They reveal that the bump is associated with a B atom crossing the hexagonal plane. This is unintuitive at first glance since the ground-state structure for BN is the non-polar layered hexagonal structure, similar to the ones shown in Figure S1. Further investigation into the fully relaxed polar structures of wurtzite-type Al$_{1-x}$B$_{x}$N alloys shows that B can form a local planar structure at high B composition, but the plane is made of two N atoms on the basal plane and one N atom along the polar axis as shown in Figure S1(c). This qualitatively explains the required extra energy for a B atom to cross the hexagonal plane since the position at the plane is not the energy local minimum. 

\begin{figure}[!h]
\centering
\includegraphics[width=0.5\linewidth]{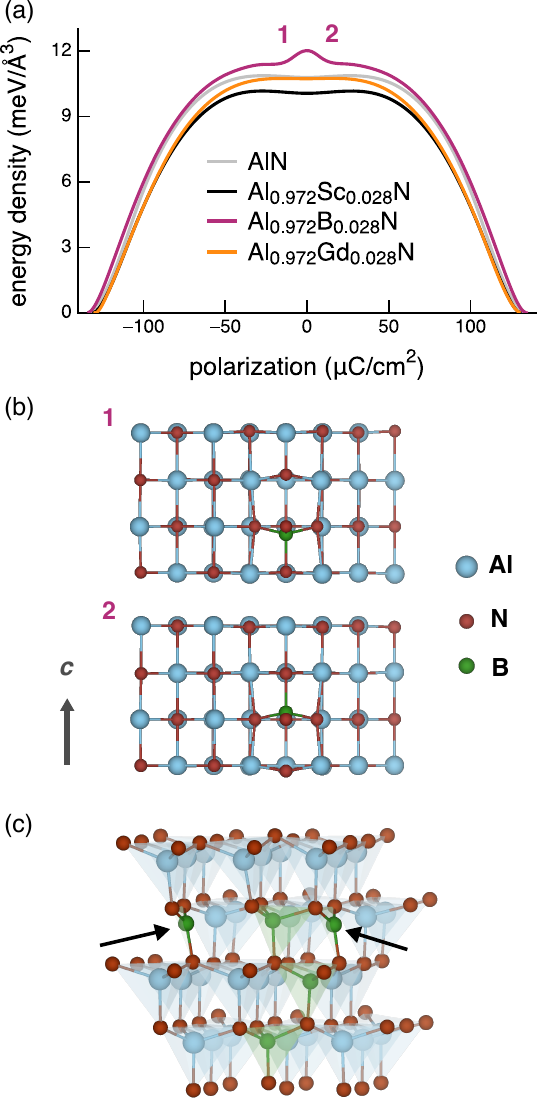}
\caption{\label{fig:mep} (a) SS-NEB-determined polarization pathways for Al$_{0.972}M_{0.028}$N ($M$ = Sc, B, Gd). (b) The structures (1) before and (2) after the peak in energy density for Al$_{0.972}$B$_{0.028}$N. In structure 1, the Al cations reach the hexagonal plane while B is still displaced downward (all tetrahedra point downward initially). The grey arrow indicates the polar axis. (c) The fully relaxed structure for one of the Al$_{0.86}$B$_{0.14}$N alloys. The two black arrows highlight the local planar structures made of B and N atoms. 
} 
\end{figure}

\clearpage
\begin{figure}[!h]
\centering
\includegraphics[width=0.6\linewidth]{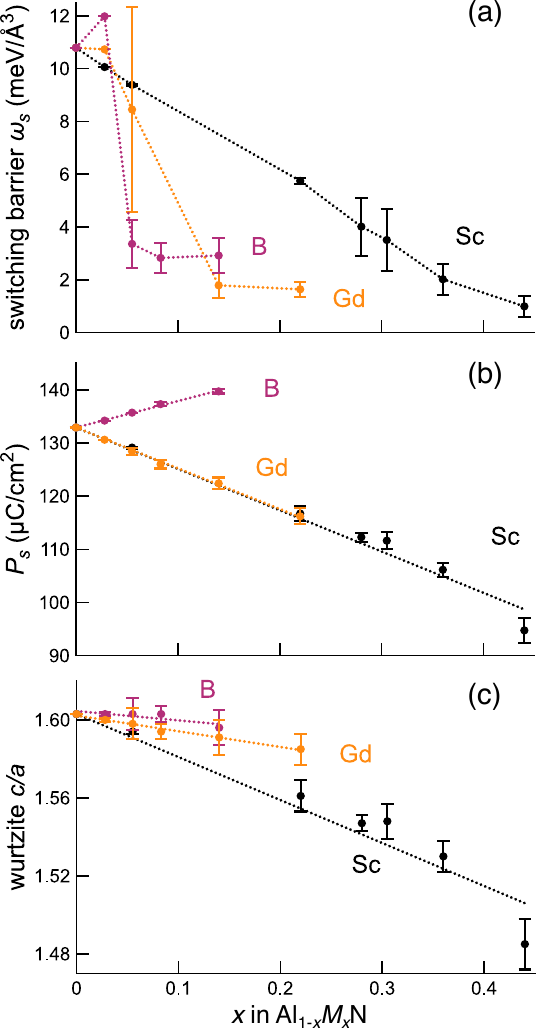}
\caption{\label{fig:sd} Predicted (a) polarization switching barriers ($\omega_s$) based on SS-NEB calculations, (b) spontaneous polarization ($P_s$), and (c) wurtzite $c/a$ lattice parameter ratio of Al$_{1-x}M_{x}$N alloys ($M$ = Sc, B, Gd). The error bars indicate the standard deviation (SD) out of 4 disordered configurations. The huge SD of the switching barrier of Al$_{0.945}$Gd$_{0.055}$N is the consequence of mixed switching mechanisms (some individual, some collective) for which switching barriers differ by $>$ 8 meV/\AA$^{3}$.
}
\end{figure}

\begin{figure}[!h]
\centering
\includegraphics[width=0.6\linewidth]{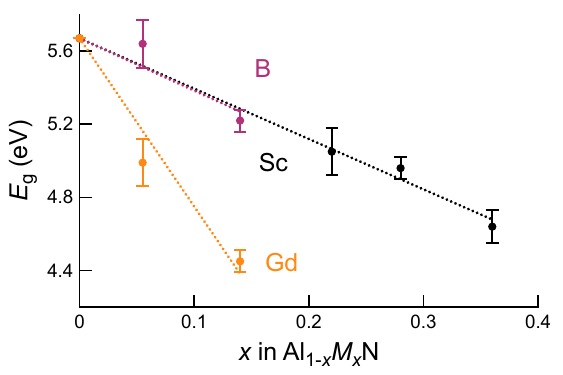}
\caption{\label{fig:gap} Predicted HSE06 band gap of Al$_{1-x}M_{x}$N alloys ($M$ = Sc, B, Gd). HSE06 slightly underestimates the band gap for pure AlN ($x$ = 0.0) by $\sim$ 0.4 eV. Dashed lines are linear fits and serve as guides for the eye.}
\end{figure}

\begin{figure}[!h]
\centering
\includegraphics[width=0.99\linewidth]{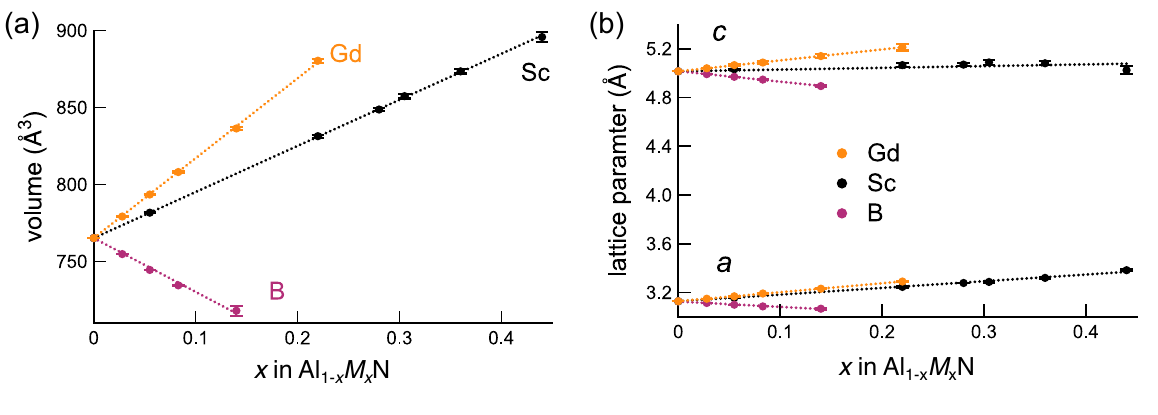}
\caption{\label{fig:vol} Predicted structural data of Al$_{1-x}M_{x}$N alloys ($M$ = Sc, B, Gd). (a) Volume of the 72-atom supercell, and (b) lattice parameters. Dashed lines show the linear fit and serve as guides for the eye.}
\end{figure}

\begin{figure}[!h]
\centering
\includegraphics[width=0.6\linewidth]{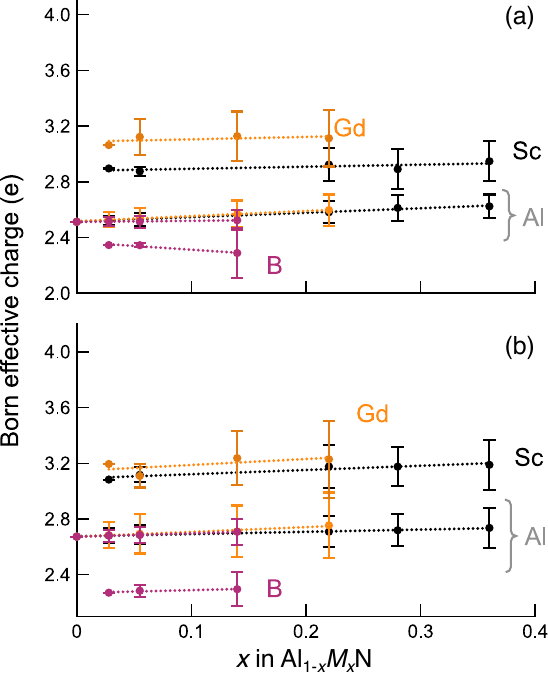}
\caption{\label{fig:bec}Born effective charges of cations in Al$_{1-x}M_{x}$N alloys ($M$ = Sc, B, Gd) as a function of alloying content (a) in the $a$--$b$ plane and (b) along the polar $c$ axis.
}
\end{figure}

\begin{figure*}[!t]
\centering
\includegraphics[width=1\linewidth]{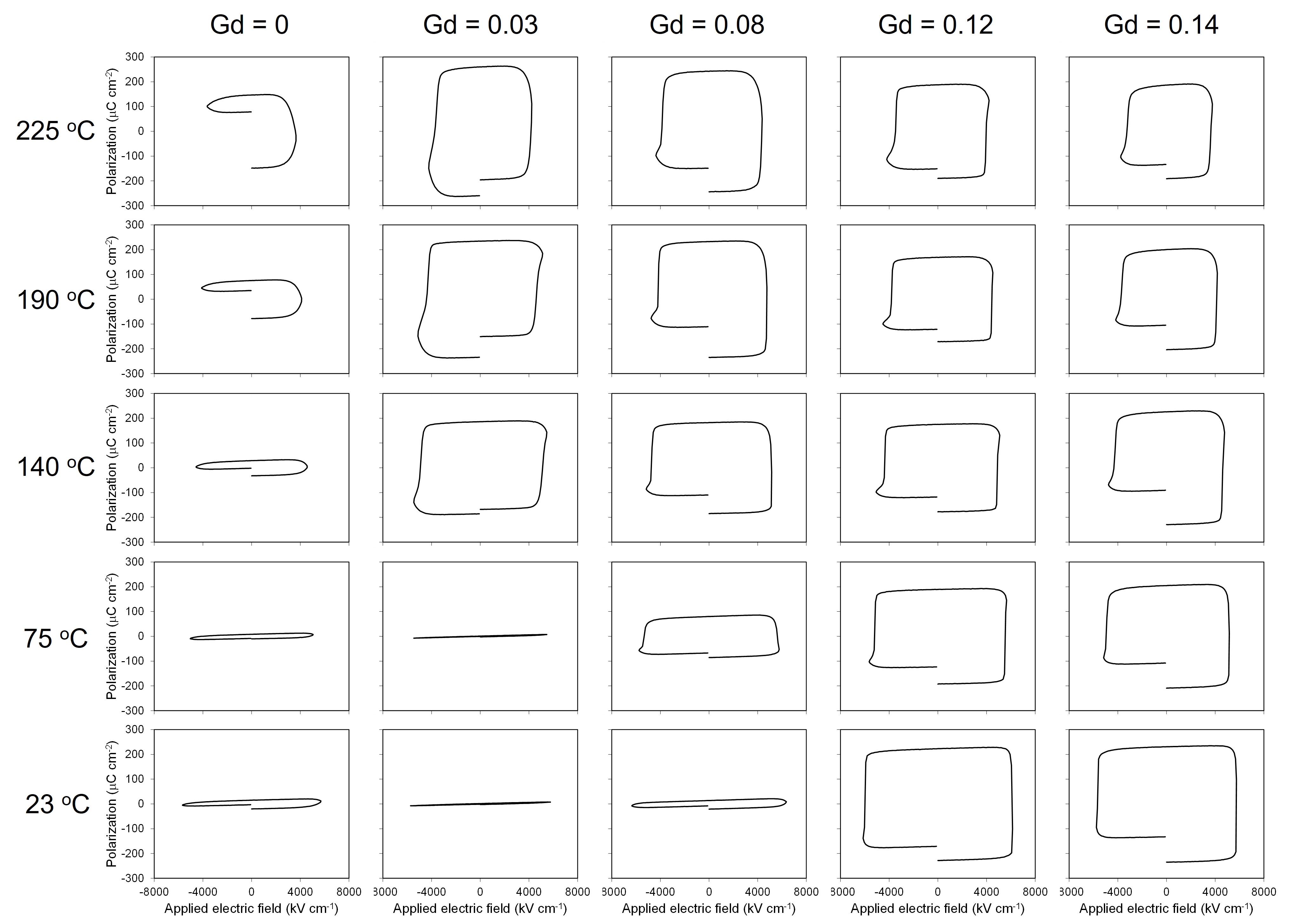}
\caption{Ferroelectric hysteresis loops for Al$_{1-x}$Gd$_{x}$N films with various Gd contents and measurement temperatures. The loops were measured with a 10 kHz triangular electric field whose amplitude is just below the breakdown field.}
\end{figure*}

\begin{figure*}[!t]
\centering
\includegraphics[width=0.9\linewidth]{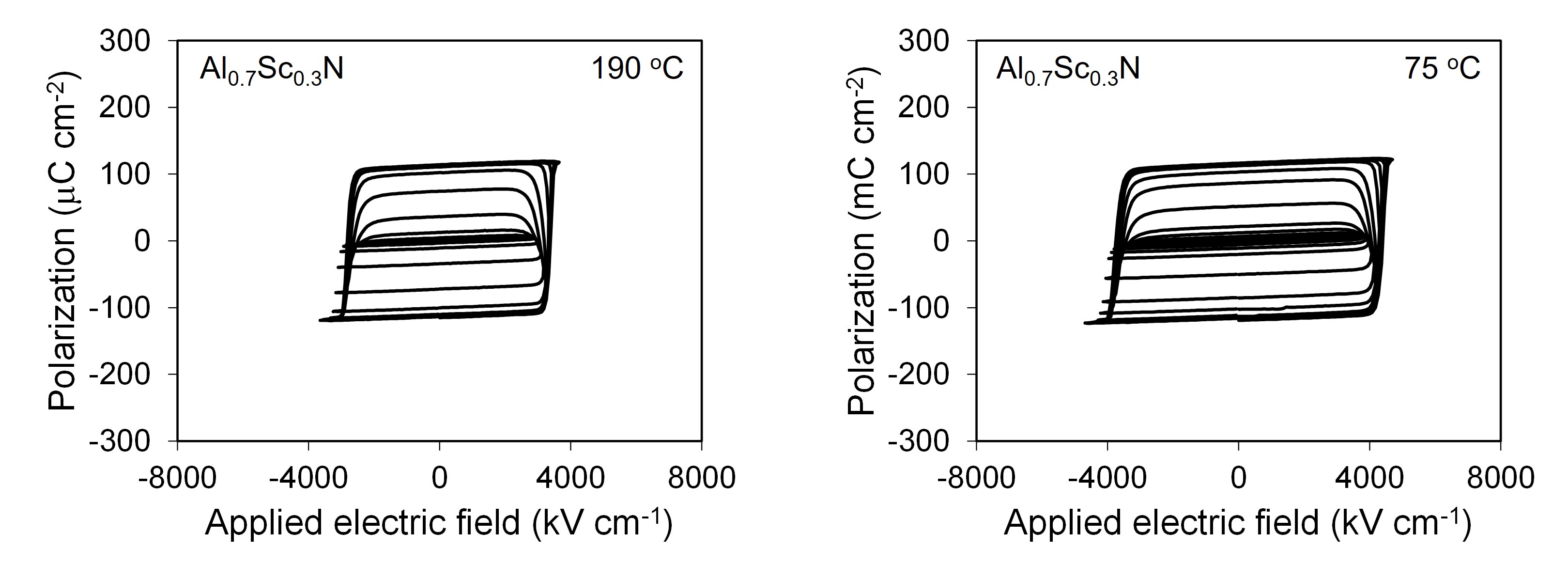}
\caption{Nested ferroelectric hysteresis loops for Al$_{0.7}$Sc$_{0.3}$N films at 190 $^{\circ}$C and 75 $^{\circ}$C.}
\end{figure*}

\begin{figure*}[!t]
\centering
\includegraphics[width=1\linewidth]{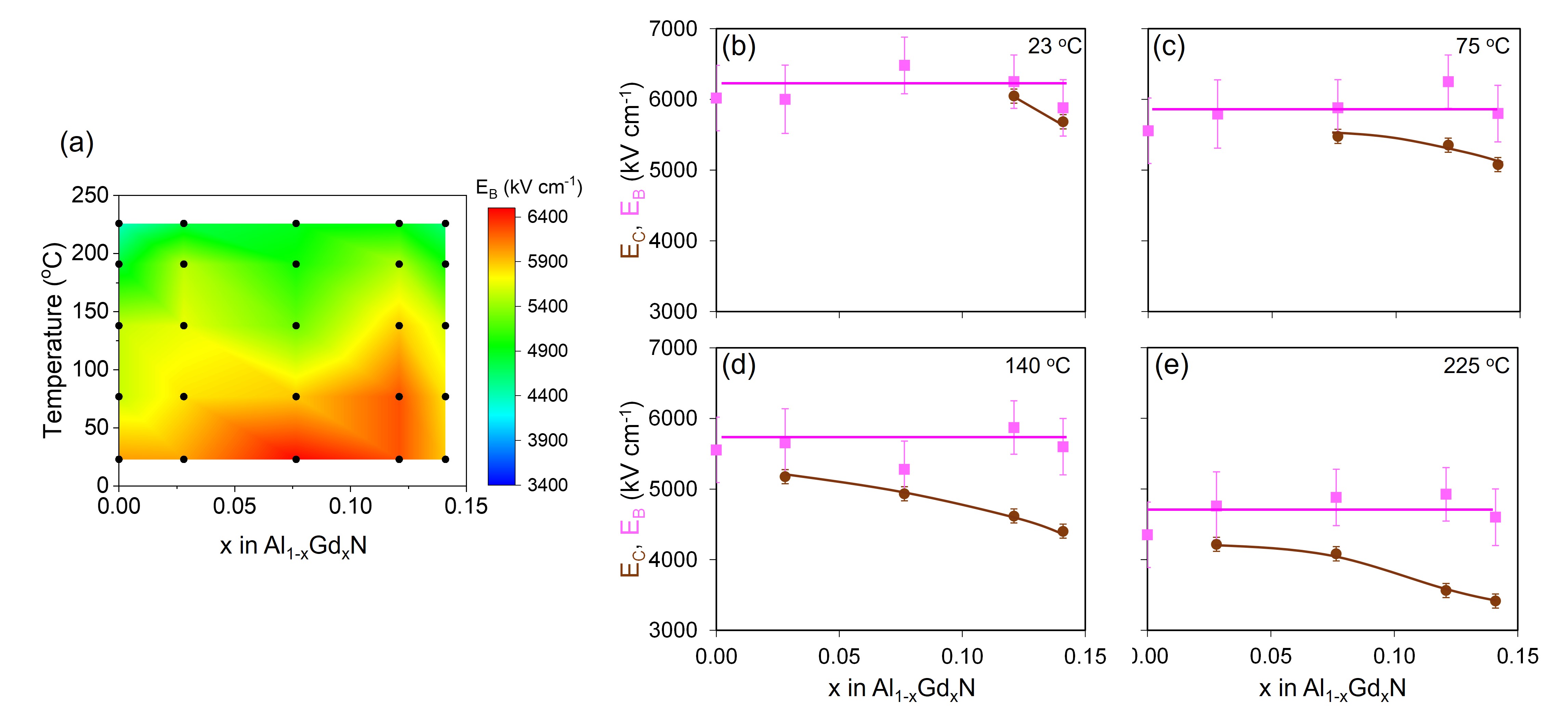}
\caption{Relationship between breakdown field and coercive field of Al$_{1-x}$Gd$_{x}$N films. (a) Breakdown field contour map showing that E$_B$ does not depend strongly on composition but does on measurement temperature. Breakdown field and coercive field comparison at (b) 23 $^{\circ}$C, (c) 75 $^{\circ}$C, (d) 140 $^{\circ}$C, and (e) 225 $^{\circ}$C. Ferroelectric switching requires $E_B > E_c$.}
\end{figure*}

\begin{figure*}[!t]
\centering
\includegraphics[width=0.5\linewidth]{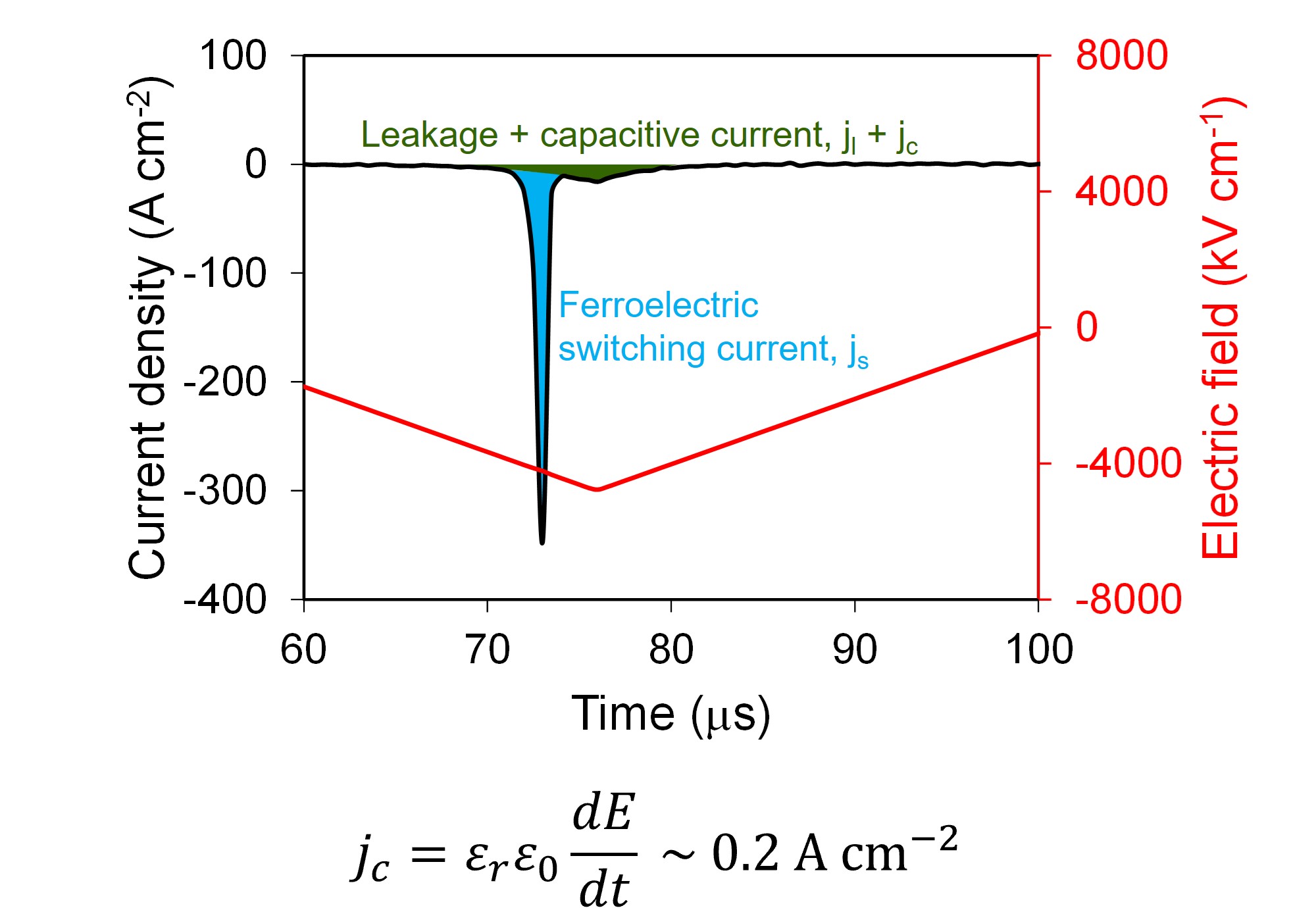}
\caption{Time dependence of measured current for remanent polarization calculation. The switching current (blue) was obtained by subtracting leakage density and capacitive current density from the total current density. The leakage current density was determined by applied field (red line). The capacitive current density was determined by permittivity and the slope of applied field, which is $\sim$0.2 A cm$^{-2}$ for Al$_{1-x}$Gd$_{x}$N films whose relative permittivity is $\sim$ 15.}
\end{figure*}

\clearpage
\newpage
\subsection{Magnetic measurements}

\begin{figure*}[]
\centering
\includegraphics[width=0.95\linewidth]{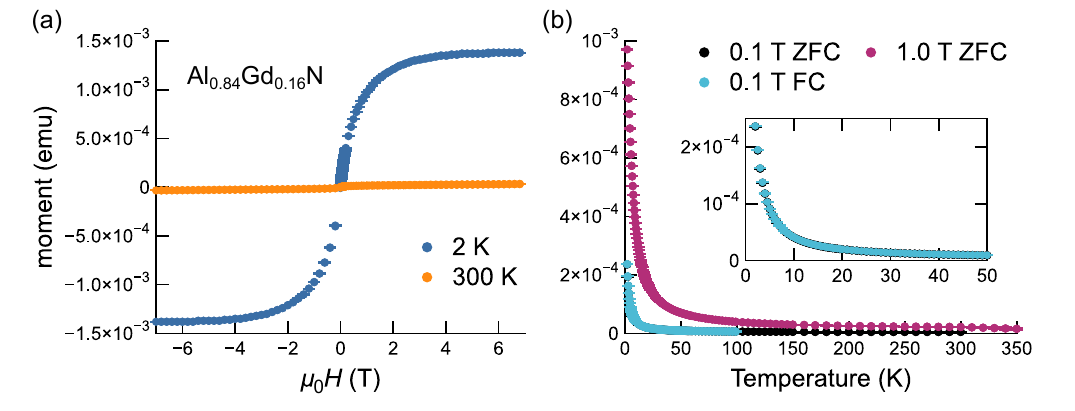}
\caption{Magnetism of Al$_{1-x}$Gd$_x$N with $x=0.16$. (a) Magnetic moment as a function of applied magnetic field at $T=2$ and $300$ K. (b) Magnetic moment as a function of temperature for several applied fields. For $\mu_0H =0.1$ T, both zero-field-cooled (ZFC) and field-cooled (FC) data are shown, and the low-temperature region is shown in the inset. Contributions from the pSi substrate have been subtracted.}
\label{fgr:mag}
\end{figure*}

Magnetic susceptibility measurements were performed on an Al$_{0.84}$Gd$_{0.16}$N film grown on a pSi substrate as a function of temperature and applied magnetic field. The substrate contribution was measured and subtracted. Figure \ref{fgr:mag}a shows the magnetic moment as a function of applied field at $T = 2$ K and $T = 300$ K. The behavior at both temperatures is consistent with paramagnetism: no hysteresis loop or net moment is observed. The temperature-dependent moment curves measured for several applied fields, shown in Figure \ref{fgr:mag}b, are similarly consistent with paramagnetism: no peak is observed. Both zero-field-cooled (ZFC) and field-cooled (FC) data were collected with applied field $\mu_0H=0.1$ T; no splitting is observed between these curves, ruling out a spin-glass ground state.

